\documentclass[pra,notitlepage,twocolumn,amsmath,amssymb,aps,superscriptaddress]{revtex4-2}

\usepackage{amsmath,amssymb,amsfonts,bbm,times}
\usepackage[dvipsnames]{xcolor}
\usepackage[T1]{fontenc}
\usepackage{braket}
\usepackage[normalem]{ulem}
\usepackage[colorlinks=true,bookmarks=false,linkcolor=RoyalBlue,urlcolor=RoyalBlue,citecolor=RoyalBlue,breaklinks]{hyperref}
\usepackage{graphicx}
\usepackage{bm}
\usepackage{physics}

\usepackage{soul}

\begin{document} 

\title{Multi-parameter quantum estimation of single- and two-mode pure Gaussian states}

\author{Gabriele Bressanini}
\affiliation{Blackett Laboratory, Imperial College London, London SW7 2AZ, United Kingdom}
\author{Marco G. Genoni}
\affiliation{Dipartimento di Fisica \textit{Aldo Pontremoli}, Universitá degli Studi di Milano, I-20133 Milano, Italy}
\author{M. S. Kim}
\affiliation{Blackett Laboratory, Imperial College London, London SW7 2AZ, United Kingdom}
\author{Matteo G. A. Paris}
\affiliation{Dipartimento di Fisica \textit{Aldo Pontremoli}, Universitá degli Studi di Milano, I-20133 Milano, Italy}

\begin{abstract}
We discuss the ultimate precision bounds on the multiparameter estimation of single- and two-mode pure Gaussian states. By leveraging on previous approaches that focused on the estimation of a complex displacement only, we derive the Holevo Cram\'er-Rao bound (HCRB) for both displacement and squeezing parameter characterizing single and two-mode squeezed states. In the single-mode scenario, we obtain an analytical bound and find that it degrades monotonically as the squeezing increases. Furthermore, we prove that heterodyne detection is nearly optimal in the large squeezing limit, but in general the  optimal measurement must include non-Gaussian resources.  On the other hand, in the two-mode setting, the HCRB improves as the squeezing parameter grows and we show that it can be attained using double-homodyne detection.
\end{abstract}
\maketitle

\section{Introduction}
Quantum devices of interest for quantum technology are complex systems characterized by many parameters. Their use in effective protocols requires
the precise characterization the underlying physical systems. This process, 
vital from both a theoretical and technological standpoint, constitutes 
the current challenge in quantum sensing and metrology. Quantum estimation theory
provides the tools to establish the ultimate limits on the precision of parameter estimation in the quantum domain, and aims to identify potential advantages with respect to classical protocols by leveraging quantum resources, including entanglement and squeezing~\cite{Giovannetti1330,Giovannetti2006,Giovannetti2011, PhysRevD.23.1693, Paris2009, Demkowicz-Dobrzanski2015a,Degen2016, Pirandola2018,Barbieri2022}. 
Multiparameter quantum metrology~\cite{MultiPerspective,Szczykulska2016,Liu_2020,DemkowiczDobrzanski2020} has received much attention in the last years, ranging from the joint estimation of unitary parameters~\cite{Genoni2013b, Bradshaw2017, Bradshaw2017a, Humphreys2013, Gagatsos2016a, Knott2016, Pezze2017,conlon2021,Conlon2023}, of unitary and loss parameters~\cite{PhysRevA.62.023815,knysh2013estimation,Vidrighin14,Altorio2015,Szczykulska_2017,Roccia_2018}, and for both spatial and time superresolution imaging~\cite{TsangPRX2016,Chrostowski2017,Rehacek2017,Rehacek18PRA,Napoli2019,Fiderer2021,Ansari2021}. From the theoretical point of view, the derivations of the ultimate bounds on the estimation precision relies on the seminal works by Helstrom~\cite{helstrom1976quantum} and Holevo \cite{Holevo2011b}; by inspecting these derivations it is immediate clear how in the quantum realm the multiparameter bounds are not a {\em trivial} generalization of the single-parameter ones, as it is indeed the case in the classical scenario. 
In fact, the potential non-commutativity of quantum-mechanical observables may lead to the incompatibility of the optimal measurements corresponding to each single parameter. 
For this reason, the standard quantum Cram\'er-Rao bound based on the symmetric logarithmic derivative (SLD) operators is in general not tight, and the departure from this bound has been indicated as a signature of {\em quantumness} of the corresponding quantum statistical model~\cite{Carollo2019,Razavian2020,Candeloro2021}. 
The Holevo Cram\'er-Rao bound (HCRB)~\cite{Holevo2011b} is a tighter bound which may differ by no more than a factor of two from the SLD Cram\'er-Rao bound~\cite{Carollo2019,TsangPRX2020} and it is in principle attainable by performing a collective measurement on an asymptotically large number of copies of the quantum state encoding the parameters. If one restricts to separable measurements on single copies, a tighter bound than the HCRB exists which is referred to as the Nagaoka-Hayashi bound (NHB)~\cite{Nagaoka2005,Hayashi,conlon2021}.
Nevertheless, it was shown that the HCRB is achievable at the single-copy level, and thus equal to the NHB, for quantum statistical models corresponding to pure states~\cite{Matsumoto_2002} and for displacement estimation tasks with Gaussian probe states~\cite{Holevo2011b}.
However, its evaluation relies on a non-trivial function minimization, which can be recast as a semidefinite program~\cite{AlbarelliPRL2019}. 
Numerical results have been presented in the context of error-corrected multiparameter quantum metrology~\cite{Gorecki2020} and for 3D magnetometry~\cite{Friel2020}. 
Analytical closed formulas for the HCRB are hard to obtain and, to the best of our knowledge, they have been derived for two-parameter estimation with pure states~\cite{Matsumoto_2002}, for generic qubit systems~\cite{Suzuki2016a}, for displacement estimation with Gaussian states~\cite{Bradshaw2017a,Bradshaw2017}, for light polarization~\cite{Jarzina2021} and for three-parameter rotations to two-qubit states~\cite{conlon2021}. 

In this work, we leverage on the methods introduced in Ref.~\cite{Bradshaw2017a} and extend the evaluation of the HCRB to the estimation of both displacement and squeezing, for single- and two-mode displaced squeezed states.
In the single-mode case, we provide an analytical expression of the HCRB for the three-parameter quantum statistical model.
Furthermore, we prove that the optimal measurement scheme must include non-Gaussian resources.
We also identify the best-performing general-dyne measurement and observe how that becomes nearly optimal in the large squeezing regime.
In the two-mode example, we resort to numerical methods for the evaluation of the HCRB and prove that the double homodyne measurement scheme is optimal in all the regimes considered, allowing us to infer the analytical form of the bound.

The manuscript is organized as follows.
In Sec.~\ref{s:gaussian} we give a brief introduction to the  Gaussian quantum state formalism for bosonic quantum systems, while in Sec.~\ref{s:qest} we introduce the basic ingredients of multiparameter quantum estimation theory.
In Sections ~\ref{s:single-mode-section} and \ref{s:two-mode-section} we present our main results, i.e. we evaluate the HCRB for single- and two-mode displaced squeezed vacuum states respectively, along with the analysis of the performance of Gaussian general-dyne measurements schemes. We conclude the manuscript in Sec.~\ref{s:conclusions} with some final remarks and outlooks.
\section{Gaussian quantum states}
\label{s:gaussian}
In this section we review key aspects of the Gaussian formalism, including Gaussian states and Gaussian measurements. We refer to the following references for a more detailed introduction~\cite{Genoni2016,serafini2017quantum,GaussFOP}. 
Let us consider a continuous variables (CV) quantum systems made up of $d$ bosonic modes described by annihilation operators $\hat{a}_j$ with $j=1,\dots,d$, that satisfy the standard bosonic commutation relations $[\hat{a}_j,\hat{a}^\dagger_k]=\delta_{jk}\hat{\mathcal{I}}$. 
Here, $\delta_{jk}$ denotes the Kronecker delta and $\hat{\mathcal{I}}$ is the identity operator. 
For each mode, one can then introduce the quadrature operators 
\begin{equation}
    \hat{q}_j=\frac{\hat{a}+\hat{a}^\dagger}{\sqrt{2}}\, ,\quad \hat{p}_j=\frac{\hat{a}-\hat{a}^\dagger}{i\sqrt{2}}
\end{equation}
that satisfy the canonical commutation relations (CCR) $[\hat{q}_j,\hat{p}_k]=i\delta_{jk}\hat{\mathcal{I}}$, where we have set $\hbar=1$.
It is convenient to arrange these operators in a vector
\begin{equation}
    \bm{\hat{r}}=(\hat{q}_1,\hat{p}_1,\dots,\hat{q}_d,\hat{p}_d)^\intercal \, ,
\end{equation}
and the CCR can now be expressed compactly as
\begin{equation}
    [\bm{\hat{r}},\bm{\hat{r}}^\intercal]=i\Omega \, ,
\end{equation}
where the symplectic matrix $\Omega$ is given by
\begin{equation}
    \Omega = \bigoplus_{j=1}^d \Omega_1 \, , \quad\quad 
    \Omega_1=
    \begin{pmatrix}
    0 & 1 \\ -1 & 0
    \end{pmatrix} \, .
\end{equation}
A generic Gaussian state $\hat{\rho}_G$ is fully characterized by its vector of first moments $\bm{\overline{r}}$ and its covariance matrix $\bm{\sigma}$, defined respectively as
\begin{equation}
\bm{\overline{r}}=\Tr[\hat{\rho}_G \bm{\hat{r}}] \, , 
\end{equation}
\begin{equation}
\bm{\sigma}=\Tr[\hat{\rho}_G \lbrace \bm{\hat{r}}-\bm{\overline{r}}, (\bm{\hat{r}}-\bm{\overline{r}})^\intercal \rbrace ] \, .
\end{equation}
Note that the conventions we use are such that the covariance matrix of a single-mode coherent state is the identity matrix, i.e. $\bm{\sigma}=\mathbb{I}_2$.
One can show that a generic $d$-mode Gaussian state can always be written as
\begin{equation}
    \hat{\rho}_G = \hat{D}^\dagger_{\bm{\overline{r}}} \hat{U}^\dagger (\bigotimes_j \hat{\nu}_j^{th}(\overline{n}_j) )\hat{U}\hat{D}_{\bm{\overline{r}}} \, ,
\end{equation}
where $\hat{D}_{\bm{\overline{r}}} = e^{i \bm{\overline{r}}^T \Omega \bm{\hat{r}}}$ is the displacement operator, $\hat{U}$ is a unitary operator generated by a purely quadratic Hamiltonian in the mode operators,
and $\hat{\nu}_j^{th}(\overline{n}_j)$ is the thermal state of a single-mode free bosonic field.
The latter reads
\begin{equation}
    \hat{\nu}_j^{th}(\overline{n}_j)= \frac{1}{1+\overline{n}_j} \left(\frac{\overline{n}_j}{1+\overline{n}_j}\right)^{\hat{a}_j^\dagger \hat{a}_j} \, ,
\end{equation}
where $\overline{n}_j$ is the mean number of bosons.
It is useful to also introduce the following parametrization of the displacement operator for the single-mode case $d=1$, namely
\begin{equation}
\hat{D}(\alpha)=e^{\alpha \hat{a}^\dagger-\alpha^* \hat{a}}=\hat{D}_{-\overline{\bm{r}}},\quad \overline{\bm{r}}=\sqrt{2}\left(\Re{\alpha},\Im{\alpha} \right)^\intercal \, .
\label{displacement_operator_1mode}
\end{equation}
By virtue of the Euler decomposition of symplectic matrices, we can express a generic single-mode Gaussian state as a displaced squeezed thermal state, i.e.
\begin{equation}
    \hat{\rho}_G = \hat{D}(\alpha)\hat{S}(\xi)\hat{\nu}^{th}(\overline{n})\hat{S}^\dagger(\xi) \hat{D}^\dagger (\alpha) \, ,
    \label{eq:singlemodeGauss}
\end{equation}
where $\hat{S}(\xi)=e^{\frac{1}{2}(\xi \hat{a}^{\dagger 2}-\xi^*\hat{a}^2)}$ is the single-mode squeezing operator and $\xi = r e^{i\chi} \in\mathbb{C}$ is the (complex) squeezing parameter.

Next, we introduce general-dyne measurements, a general class of Gaussian measurements whose corresponding positive-operator-valued measure (POVM) originates from a generalization of the well known over-completeness relation of coherent states~\cite{Genoni2016,serafini2017quantum}, namely
\begin{equation}
\frac{1}{(2\pi)^d}\int d^{2d}\bm{r}_m\, \hat{D}_{\bm{r}_m}^\dagger \,\hat{\rho}_m  \,\hat{D}_{\bm{r}_m}=\hat{\mathcal{I}} \, ,
\end{equation}
where $\hat{\rho}_m$ is a generic $d-$mode Gaussian state with null vector of first moments and covariance matrix $\bm{\sigma}_m$. A general-dyne detection is said to be \textit{ideal} if $\hat{\rho}_m$ is a pure state. 
The POVM associated with this measurement thus reads
\begin{equation}
    \hat{\Pi}_{\bm{r}_m}=\frac{ \hat{D}_{\bm{r}_m}^\dagger \,\hat{\rho}_m  \,\hat{D}_{\bm{r}_m}}{(2\pi)^d} \, .
\end{equation}
Hence, $\bm{\sigma}_m$ defines the specific measurement scheme, while $\bm{r}_m$ represents the measurement outcome. 
The homodyne and heterodyne detections are easily recovered in this formalism. For example, in a single-mode scenario, the covariance matrix associated with the $\hat{q}$ quadrature measurement is
\begin{equation}
\bm{\sigma}_m = \lim_{s\rightarrow -\infty}\mqty(e^{2s} & 0 \\ 0 & e^{-2s}) = \lim_{z\rightarrow 0}\mqty(z & 0 \\ 0 & \frac{1}{z}),
\end{equation}
while the heterodyne detection is retrieved with the substitution $\bm{\sigma}_m=\mathbb{I}_2$ (the POVM elements are projectors on coherent states).
Lastly, the probability distribution of outcomes of a general-dyne detection on a Gaussian input state $\hat{\rho}_G$ characterized by vector of first moments $\overline{\bm{r}}$ and covariance matrix $\bm{\sigma}$ 
{corresponds to a multi-variate Gaussian distribution centered in $\overline{\bm{r}}$ and with covariance matrix $\boldsymbol{\Sigma}=(\bm{\sigma} +\bm{\sigma}_m)/2$.}

\section{Multi-parameter Quantum Estimation Theory}
\label{s:qest}
Let us consider a quantum statistical model $\hat{\rho}_{\bm{\theta}}$, i.e. a family of quantum states labelled by a vector of $d$ real parameters $\bm{\theta}=(\theta_1,\dots,\theta_d)^\intercal$. We want to estimate the value of $\bm{\theta}$ from the outcomes of $M$ independent measurements $\bm{x}=(x_1,\dots,x_M)^\intercal$ described by the POVM $\hat{\Pi}_x$, using a suitable unbiased estimator $\tilde{\bm{\theta}}(\bm{x})$.
The accuracy of the latter may be evaluated in terms of its covariance matrix, defined as 
\begin{equation}
\bm{V}(\tilde{\bm{\theta}}) =\int d\bm{x}\, p(\bm{x}\vert\bm{\theta})(\tilde{\bm{\theta}}(\bm{x})-\bm{\theta})(\tilde{\bm{\theta}}(\bm{x})-\bm{\theta})^\intercal \, .
\end{equation}
Here, $p(\bm{x}\vert\bm{\theta})=\Pi_{j=1}^M p(x_j\vert\bm{\theta})=\Pi_{j=1}^M \Tr[\Pi_x \rho_{\bm{\theta}}]$, where we have implicitly assumed that the $M$ measurements are independent of each other.
The covariance matrix satisfies the Cram\'er-Rao bound (CRB)
\begin{equation}
\label{classicalCRB}
    \bm{V}(\tilde{\bm{\theta}}) \geq \frac{1}{M} \bm{F}^{-1} \, ,
\end{equation}
where $\bm{F}$ is the Fisher Information (FI) matrix, whose elements are defined as
\begin{equation}
\bm{F}_{\mu\nu} = \int d\bm{x}\, p(\bm{x}\vert\bm{\theta}) \left( \partial_\mu \log{p(\bm{x}\vert\bm{\theta})}\right) \left( \partial_\nu \log{p(\bm{x}\vert\bm{\theta})}\right) \, .
\label{FImatrix}
\end{equation}
Throughout this work we will use the notation $\partial_\mu \equiv \pdv{\theta_\mu}$. 
In particular, for a Gaussian probability distribution $p({\bm x}|\bm{\theta})$ centered in $\overline{\bm x}$ and with covariance matrix $\boldsymbol{\Sigma}$, one has
\begin{equation}
\bm{F}_{\mu\nu} = (\partial_\mu \overline{\bm x}^{\intercal}) \boldsymbol{\Sigma}^{-1} (\partial_\nu \overline{\bm x}) + \frac12 \hbox{Tr}\left[
\boldsymbol{\Sigma}^{-1}( \partial_\mu \boldsymbol{\Sigma}) \boldsymbol{\Sigma}^{-1} (\partial_\nu \boldsymbol{\Sigma} )
\right] \,.
\label{FImatrixGauss}
\end{equation}
The CRB is attainable, at least asymptotically, by choosing a suitable efficient estimator.
Furthermore, quantum mechanics allows us to find tighter precision bounds, which only depend on the statistical model.
Let us introduce the symmetric logarithmic derivative (SLD) operators, implicitly defined by the following Lyapunov equation
\begin{equation}
\partial_\mu \hat{\rho}_{\bm{\theta}} = \frac{\hat{L}_\mu \hat{\rho}_{\bm{\theta}} +\hat{\rho}_{\bm{\theta}} \hat{L}_\mu }{2}\, .
\label{SLDs}
\end{equation}
For pure statistical models $\hat{\rho}_{\bm{\theta}}=\ketbra{\psi_{\bm{\theta}}}$ one can easily solve the equation above and show that the SLD assumes the following simple form:
\begin{equation}
\hat{L}_\mu = 2\partial_\mu\hat{\rho}_{\bm{\theta}} = 2(\ket{\psi_{\bm{\theta}}}\!\!\bra{\partial_\mu\psi_{\bm{\theta}}}+\ket{\partial_\mu\psi_{\bm{\theta}}}\!\!\bra{\psi_{\bm{\theta}}}).
\label{SLDpure}
\end{equation}
We then use the SLDs to define  the Quantum Fisher Information (QFI) matrix
\begin{equation}
\bm{Q}_{\mu\nu} = \Tr[\hat{\rho}_{\bm{\theta}}\frac{\hat{L}_\mu \hat{L}_\nu + \hat{L}_\nu \hat{L}_\mu}{2} ] \, ,
\label{QFImatrix}
\end{equation}
which is in turn used to derive the matrix quantum Cram\'er-Rao bound ~\cite{helstrom1976quantum}
\begin{equation}
\bm{V}(\tilde{\bm{\theta}})\geq\frac{1}{M} \bm{Q}^{-1}\, .
\label{quantum_CRB_matrix}
\end{equation}
In the single-parameter scenario, the Cram\'er-Rao inequality Eq.~\eqref{quantum_CRB_matrix} is a scalar bound that can be attained by projective measurement of the (Hermitian) SLD operator.
On the other hand, in the multi-parametric setting the quantum  Cram\'er-Rao bound is in general not tight, because of the possible non-commutativity of the SLD operators. 
As it is usually more convenient to deal with scalar bounds, we use a real, positive, $d\times d$ weight matrix $\bm{W}$ to introduce the following scalar inequality, which we refer to as the SLD Cram\'er-Rao bound (SLD-CRB):
\begin{equation}
\Tr[\bm{W}\bm{V}]\geq \Tr[\bm{W}\bm{Q}^{-1}]\equiv C^S(\bm{\theta},\bm{W})\, .
\label{scalar_CRB}
\end{equation}
Different $\bm{W}$ matrices are used to weigh the uncertainties of the parameters differently. In this paper we will set $\bm{W}=\mathbb{I}_d$, so that the inequality above bounds $\Tr[\bm{V}]$, i.e. the sum of the variances of each parameter's estimate. Like the corresponding matrix bound, the SLD-CRB in Eq.~\eqref{scalar_CRB} is also in general not attainable. We can also define the \textit{most informative} bound as a minimization of the classical bound over all possible quantum measurements, i.e.
\begin{equation}
C^{\textrm{MI}}(\bm{\theta}) = \min_{\textrm{POVM}} \Tr[\bm{F}^{-1}] \, ,
\end{equation}
which is in general larger than the SLD-CRB. However, there exists a tighter bound, known as the HCRB, such that the following chain of inequalities holds
\begin{equation}
\Tr[\bm{V}]\geq C^{\textrm{MI}} (\bm{\theta}) \geq C^H(\bm{\theta}) \geq C^S (\bm{\theta})\, .
\label{holevo_bound}
\end{equation}
The HCRB $C^H (\bm{\theta})$ is defined via the following minimization~\cite{Holevo2011b}
\begin{equation}
C^H (\bm{\theta}) = \min_{\hat{\bm{X}}} h_{\bm{\theta}}\left[\hat{\bm{X}}\right] \, ,
\label{holevo_def}
\end{equation}
where $\hat{\bm{X}}=(\hat{X}_1,\dotsc,\hat{X}_d)$ is a vector of Hermitian operators satisfying the locally unbiased conditions
\begin{align}
\Tr[\hat{\rho}_{\bm{\theta}} \hat{X}_j] &= 0 \,,
\label{locally_unbiased_cond1} \\
\Tr[(\partial_j{\hat{\rho}_{\bm{\theta}}})\hat{X}_k] &=\delta_{jk} \, . 
\label{locally_unbiased_cond2}
\end{align}
Finally, the function to minimize reads
\begin{equation}
h_{\bm{\theta}} [\hat{\bm{X}}]= \Tr[\Re{\bm{Z}_{\bm{\theta}}[\hat{\bm{X}}]}] + \norm{\Im{\bm{Z}_{\bm{\theta}}[\hat{\bm{X}}]}}_1 \, ,
\label{h_function}
\end{equation}
where $\norm{A}_1 \equiv \Tr[\sqrt{A^\dagger A}]$  and $\bm{Z}_{\bm{\theta}}[\hat{\bm{X}}]$ is a matrix of operators defined as
\begin{equation}
\bm{Z}_{\bm{\theta}}[\hat{\bm{X}}]_{jk} = \Tr[\hat{\rho}_{\bm{\theta}}\hat{X}_j \hat{X}_k]\, .
\label{Z_matrix_elements}
\end{equation}
The HCRB is typically regarded as the most fundamental scalar bound in multi-parameter quantum estimation theory, as it can be shown to be attainable by performing a collective measurement on an asymptotically large number of copies of the quantum state $\hat{\rho}_{\bm{\theta}}$ encoding the parameters. 
Nevertheless, it was shown that the HCRB is actually attainable at the single copy level for displacement estimation tasks of Gaussian states and for quantum statistical models encoded in pure quantum states.
The quantity $C^H(\bm{\theta})$ may also be  bounded ~\cite{Carollo2019,TsangPRX2020} as follows
\begin{equation}
C^S(\bm{\theta}) \leq C^H (\bm{\theta}) \leq (1+R) C^S (\bm{\theta}) \leq 2 C^S (\bm{\theta})\, ,
\label{eq:bounds}
\end{equation}
where
\begin{align}
R=\norm{i\bm{Q}^{-1}(\bm{\theta})\bm{D}(\bm{\theta})}_\infty
\label{eq:Rquantumness}
\end{align}
has been referred to as the asymptotic incompatibility (AI) of the corresponding quantum statistical model \cite{Razavian2020,Candeloro2021}. In the formula above, $\norm{\bm{A}}_\infty$ denotes the largest eigenvalue in modulus of $\bm{A}$, and $\bm{D}(\bm{\theta})$ is known as the Uhlmann curvature, whose matrix elements are defined as
\begin{equation}
\bm{D}_{\mu\nu} = -\frac{i}{2}\Tr[\hat{\rho}_{\bm{\theta}}[\hat{L}_\mu,\hat{L}_\nu]]\, .
\label{Uhlmann}
\end{equation}
The AI measure provides a bound for the true normalized gap between the HCRB and the SLD-CRB bounds, as follows
\begin{align}
\frac{
C^H(\boldsymbol{\theta}) -
C^S(\boldsymbol{\theta})
}{C^S(\boldsymbol{\theta})} \leq R\,.
\end{align}
One can also prove \cite{Carollo2019} 
that $0\leq R\leq 1$, hence the SLD-CRB gives an 
estimate of the HCRB up to a factor two. As it is also clear from Eq.~\eqref{eq:bounds}  and Eq.~\eqref{Uhlmann}, the HCRB and the SLD-CRB coincide whenever the Uhlmann curvature matrix elements are all equal to zero, that is whenever the average values of the commutators of the SLD operators on the quantum statistical model $\hat{\rho}_{\bm{\theta}}$ are equal to zero.

In the following Sections, we apply the tools of quantum estimation theory to fully characterize general classes of single-mode and two-mode displaced squeezed vacuum states, which correspond to three-parameter quantum statistical models. The choice of focusing on pure statistical models originates from the two following observations. 
First, from an experimental point of view, the description of optical frequencies radiation modes at temperatures accessible in a lab with the vacuum state is a very good approximation. 
Furthermore, we recall that the HCRB is defined as a minimization problem over a set of peculiar Hermitian matrices, however this difficulty is eased by the particular structure of the quantum statistical models considered and we are thus able to compute the bound analytically, or at least to greatly simplify its numerical evaluation.

\section{Multi-parameter quantum estimation of single mode pure Gaussian state}
\label{s:single-mode-section}
Let us consider a quantum statistical model given by single-mode displaced squeezed vacuum states
\begin{equation}
\ket{\psi_{\bm{\theta}}} = \hat{D}(\alpha)\hat{S}(r)\ket{0} \, ,
\label{stat_model_1}
\end{equation}
where $\alpha$ is a complex displacement and $r\geq 0$ is the (real) squeezing parameter. 
As already mentioned, Eq.~\eqref{stat_model_1} represents the most general single-mode pure Gaussian state (up to a phase shift), corresponding to Eq.~\eqref{eq:singlemodeGauss} with $\overline{n} = 0$.
We want to assess the ultimate bound to the joint estimation precision of the three real parameters that characterize the statistical model, namely
\begin{equation}
    \bm{\theta} = (\Re{\alpha},\Im{\alpha},r)^\intercal = (\theta_1,\theta_2,\theta_3 )^\intercal \, .
    \label{vector_of_parameters}
\end{equation}
In the following, we compute the SLD-CRB, the HCRB and discuss the experimental attainability of the latter.
In order to simplify the notation we introduce the following Hilbert space basis
\begin{equation}
\ket{e_n} = \hat{D}(\alpha)\hat{S}(r)\ket{n} \, ,
\end{equation}
where $\ket{n}$ is the $n-$boson Fock state. Hence, the quantum statistical model simply reads $\ket{\psi_{\bm{\theta}}} = \ket{e_0}$.
In Appendix \ref{a:single-mode-calculations}, we outline the detailed calculation of the SLDs and the QFI matrix. We find the latter to be
\begin{equation}
\bm{Q} = \mqty (4e^{-2r} & 0  & 0 \\ 0 & 4e^{2r} & 0  \\ 0 & 0 & 2) \, .
\end{equation}
As expected due to symmetry reasons, $\bm{Q}$ does not depend on the displacement $\alpha$. 
Additionally, the diagonal matrix elements of reveal that, at the single parameter estimation level, squeezing does not affect the estimation of $r$ itself, however it plays a role in the estimation of the displacement.
We can understand this intuitively: squeezing decreases the variance of the $\hat{p}$ quadrature and increases that of the $\hat{q}$ quadrature. Consequently, even before computing the CRB, we may conclude that squeezing is not a useful resource for the simultaneous estimation of the three parameters: as $r$ grows, the estimation error of $\theta_1$ decreases, eventually reaching zero, while the variance of $\theta_2$ grows indefinitely. This simple argument tells us that $r$ is not a useful resource already in a displacement-only estimation scenario. We can now compute the SLD-CRB, namely
\begin{equation}
\Tr[\bm{V}^{-1}]\geq  \Tr[\bm{Q}^{-1}] = C^S=\frac{1+\cosh{(2r)}}{2} \, .
\label{crb_singlemode}
\end{equation}
We also compute the quantumness $R$ of our statistical model.
In particular, the Uhlmann curvature $\bm{D}$ reads
\begin{equation}
\bm{D} = \mqty (0 & 4 & 0 \\ -4 & 0 & 0 \\ 0 & 0 & 0) \, ,
\end{equation}
hence, via Eq.~(\ref{eq:Rquantumness}), we obtain $R=1$, i.e. the maximum degree of incompatibility between the parameters. 
We remark that the same result would be obtained for the estimation of the displacement parameters only, and the maximum incompatibility can be ascribed to the incompatibility of the $\hat{q}$ and $\hat{p}$ quadrature operators. 

\begin{figure}
\centering 
\includegraphics[scale=0.41]{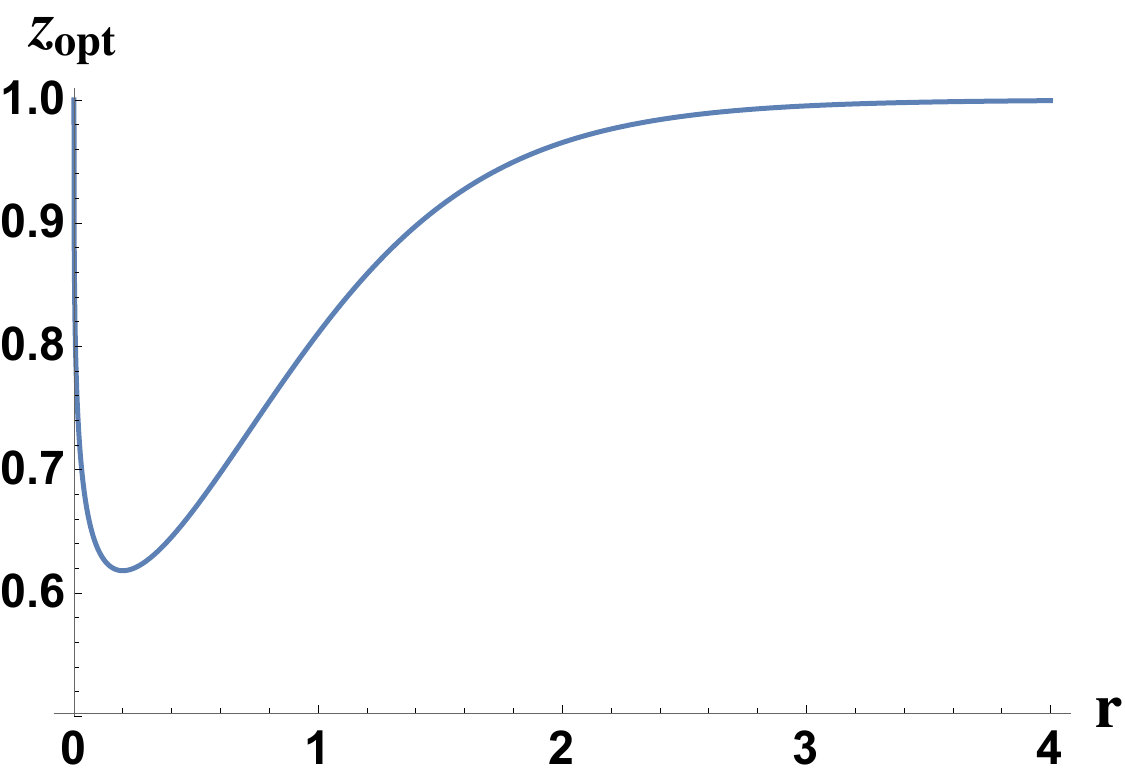}
\caption{
The implicit function $z_{opt}(r)$, obtained from numerically minimizing Eq.~\eqref{function_difference} at fixed value of the squeezing parameter $r$. We remind that $z_{opt}$ determines the optimal measurement - among general-dyne detection schemes - for our three-parameter estimation problem and that $z_{opt}=1$ corresponds to heterodyne measurement.}
\label{optimal_z}
\end{figure}

\begin{figure}
\centering 
\includegraphics[scale=0.41]{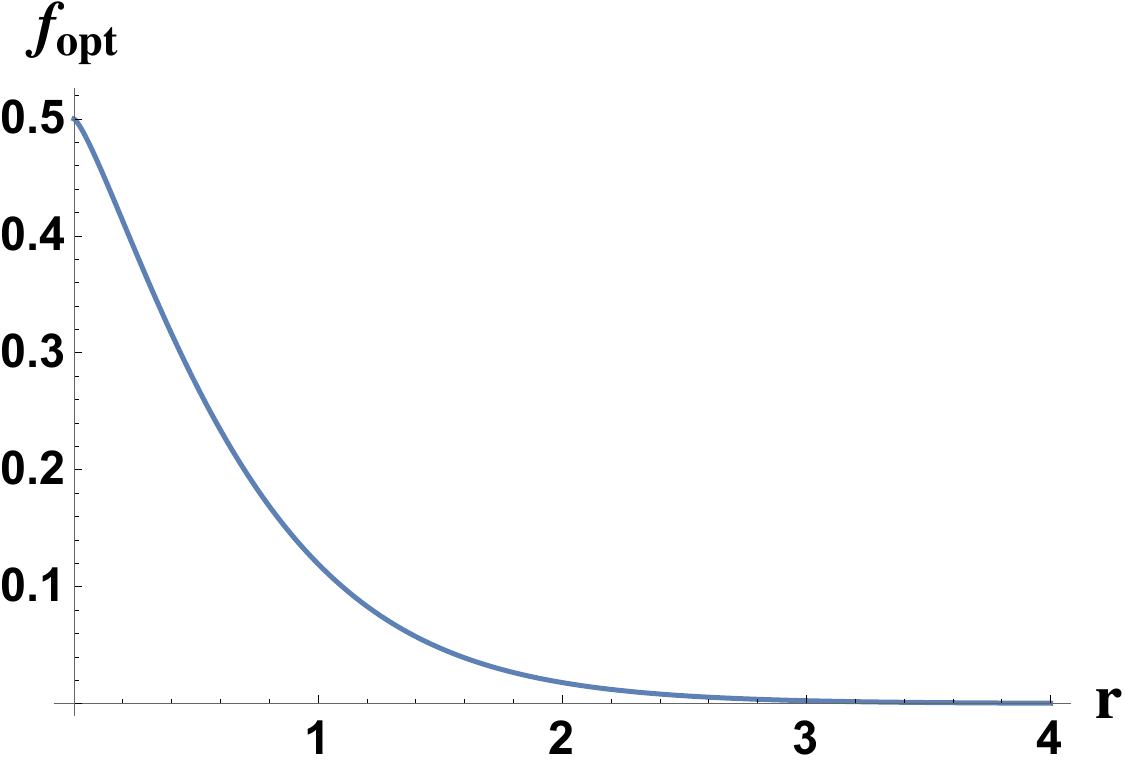}
\caption{The function $f_{opt}(r)=f(z_{opt}(r),r)$, representing the difference between the optimal estimation precision obtained with a POVM belonging to the general-dyne class and the HCRB. The optimization is carried out at fixed $r$.
We notice that the function is monotonically decreasing in $r$, hence the estimation performance achievable by the optimal general-dyne matches the HCRB only for large values of squeezing.}
\label{optimal_f}
\end{figure}

We are now ready to tackle the calculation of the HCRB. The derivatives of the quantum statistical model with respect to the three parameters can be found in the Appendix~\ref{a:single-mode-calculations} (see Eqs.~(\ref{d1}), (\ref{d2}) and (\ref{d3}) ). 
We recall that the HCRB is defined as a minimization of the function $h_{\bm{\theta}}[\hat{\bm{X}}]$ in Eq.~\eqref{h_function} over all possible vectors of Hermitian operators $\hat{\bm{X}}$ satisfying the locally unbiased conditions Eq.\eqref{locally_unbiased_cond1} and Eq.\eqref{locally_unbiased_cond2}.
In our case $\hat{\bm{X}}=\lbrace \hat{X}_1,\hat{X}_2,\hat{X}_3\rbrace$, and the constraint imposed by Eq.~\eqref{locally_unbiased_cond1} implies
\begin{equation}
\bra{e_0} \hat{X}_1 \ket{e_0}=\bra{e_0} \hat{X}_2 \ket{e_0}=\bra{e_0} \hat{X}_3 \ket{e_0} = 0\, .
\end{equation}
Since our quantum statistical model is pure, 
Eq.~\eqref{locally_unbiased_cond2} can be expressed as 
\begin{equation}
\delta_{jk}=\Tr[(\partial_j {\hat{\rho}_{\bm{\theta}}})\hat{X}_k]= \bra{\partial_j\psi_{\bm{\theta}}} \hat{X}_k \ket{\psi_{\bm{\theta}}} + \bra{\psi_{\bm{\theta}}} \hat{X}_k \ket{\partial_j\psi_{\bm{\theta}}} \, .
\label{locally_unbiased_condition2_v2}
\end{equation}
The latter imposes three additional conditions for each of our Hermitian operators. As an example, we find that $\hat{X}_1$ must satisfy
\begin{align}
1 & =\Tr[(\partial_1{\hat{\rho}_{\bm{\theta}}}) \hat{X}_1] \Rightarrow \Re{\bra{e_0}\hat{X}_1\ket{e_1}}=\frac{e^r}{2} \, ,  \\
0 & =\Tr[(\partial_2 {\hat{\rho}_{\bm{\theta}}}) \hat{X}_1] \Rightarrow \Im{\bra{e_0}\hat{X}_1\ket{e_1}}=0 \, , \\
0 &=\Tr[(\partial_3{\hat{\rho}_{\bm{\theta}}}) \hat{X}_1] \Rightarrow \Re{\bra{e_0}\hat{X}_1\ket{e_2}}=0 \, .
\end{align}
These in turn imply 
\begin{equation}
{\bra{e_0}\hat{X}_1\ket{e_1}} = \frac{e^r}{2} \, , \quad {\bra{e_0}\hat{X}_1\ket{e_2}} =i\beta \, ,
\end{equation}
where $\beta\in\mathbb{R}$. 
Analogous calculations yield the following conditions on the matrix elements of $\hat{X}_2$ and $\hat{X}_3$: 
%
\begin{align}
{\bra{e_0}\hat{X}_2\ket{e_1}} &= -\frac{ie^{-r}}{2} \,, \quad &{\bra{e_0}\hat{X}_2\ket{e_2}} &= i\gamma \,,\\
{\bra{e_0}\hat{X}_3\ket{e_1}} &= 0 \,,  \quad &{\bra{e_0}\hat{X}_3\ket{e_2}} &=\frac{1}{\sqrt{2}}+i\delta \, ,
\end{align}
where $\gamma$ and $\delta$ are free real parameters. 
Clearly, the hermiticity of $\hat{X}_j$ imposes additional constraints on the matrix elements of these operators.
Now comes the great simplification of dealing with pure states statistical models: in Ref.~\cite{Matsumoto_2002} it is shown that we can set to zero all the matrix elements of $\hat{X}_j$ not involved in the aforementioned constraints. This is equivalent to saying that it is sufficient to study these Hermitian operators in the Hilbert subspace spanned by the partial derivatives of the statistical model.
After some algebra, we obtain the matrix
\begin{equation}
\bm{Z}_{\bm{\theta}}[\hat{\bm{X}}]=\mqty( \frac{e^{2r}}{4}+\beta^2 & \frac{i}{4} +\beta\gamma & \frac{i\beta}{\sqrt{2}}+\beta\delta \\ 
-\frac{i}{4}+\beta\gamma & \frac{e^{-2r}}{4}+\gamma^2 & \frac{i\gamma}{\sqrt{2}}+\gamma\delta \\
	-\frac{i\beta}{\sqrt{2}} &  -	\frac{i\gamma}{\sqrt{2}}+\gamma\delta & \frac{1}{2}+\delta^2) \, .
\end{equation}
We can then use the definition in Eq.~\eqref{h_function} to compute the function $h_{\bm{\theta}}[\hat{\bm{X}}]$, namely
\begin{equation}
h_{\bm{\theta}}[\hat{\bm{X}}] = \frac{\cosh{(2r)}+1}{2}+\beta^2+\gamma^2+\delta^2 +\frac{1}{2}\sqrt{1+8\beta^2+8\gamma^2} \, ,
\end{equation}
and perform a minimization over the real parameters $\beta,\gamma$ and $\delta$.
The optimization is trivial and yields the HCRB, i.e.
\begin{equation}
C^H = \frac{\cosh{(2r)}+2}{2} \, .
\label{holevo_bound_1_mode}
\end{equation}
As we can see, optimal performance is achieved for $r=0$ and we also notice that $C^H > C^S \,\, \forall r$, thus confirming that the SLD-CRB is in general not tight. The latter coincides with $C^H$ only in the asymptotic limit $r\rightarrow\infty$. This also means that in this case the AI measure $R$ overestimates the true gap between $C^H$ and $C^S$.
As mentioned above, it has been proven in Ref.~\cite{Matsumoto_2002} that the HCRB is achievable via single-copy measurements for pure quantum statistical models.  However, we do not know in general what measurement saturates said bound.
In Ref.~\cite{Genoni2013b} it was shown that if we are interested in estimating the complex displacement $\alpha$ only, then the optimal measurement corresponds to heterodyne detection.
As we now want to evaluate the the ultimate bound including the squeezing parameter, we study the more general class of general-dyne detections, and compare their performances to the HCRB. 
We remind the reader that the covariance matrix $\bm{\sigma}_m$ which characterizes the specific general-dyne measurement scheme is that of a generic Gaussian state with zero displacement. As we are looking for the optimal measurement, we also restrict ourselves to ideal (noiseless) measurements, thus corresponding to pure Gaussian states.
This means that the measurement is fully described by a covariance matrix of the form
\begin{equation}
    \bm{\sigma}_m = \bm{S} \bm{S}^\intercal \, ,
\end{equation}
where $\bm{S}$ is a symplectic matrix corresponding to a unitary evolution induced by a purely quadratic Hamiltonian in the quadrature operators. 
The Euler decomposition theorem states that any symplectic matrix can be expressed in terms of symplectic matrices corresponding to phase rotations, beam splitters and local squeezing only. 
However, since we are dealing with single-mode statistical models, we do not need to include beam splitters and it is easy to convince ourselves that phase rotations do not play a role (we have also verified this intuition by numerical means), hence we may discard those as well.
As a result, $\bm{S}$ is just the usual symplectic matrix that corresponds to single-mode squeezing, hence $\bm{\sigma}_m$ reads
\begin{equation}
\bm{\sigma}_m = \mqty (z & 0 \\ 0 & z^{-1}) \, ,
\end{equation}
with $z>0$. 
The covariance matrix of the statistical model is 
\begin{equation}
\bm{\sigma}=\mqty (e^{2r} & 0 \\ 0 & e^{-2r}) \, ,
\end{equation}
and the conditional probability distribution of outcomes has Gaussian form, centered in $\overline{\bm x}= (\theta_1,\theta_2)^\intercal$ and with covariance matrix given by $\boldsymbol{\Sigma}=(\boldsymbol{\sigma}+\boldsymbol{\sigma}_m)/2$.
The FI matrix can then be obtained via Eq.~(\ref{FImatrixGauss}), yielding the following non-zero matrix elements:
\begin{align}
    \bm{F}_{\textrm{gen},11} &=\frac{4}{e^{2r}+ z} \, , 
    \label{fel1} \\
    \bm{F}_{\textrm{gen},22} &=\frac{4}{e^{-2r}+ z^{-1}} \, , 
    \label{fel2} \\
    \bm{F}_{\textrm{gen},33} &=\frac{2(e^{4r}+z^2)}{(e^{2r}+ z)^2} \, .
        \label{fel3} 
\end{align}
The bound on the estimation precision is thus given by
\begin{equation}
\Tr[\bm{F}_{\textrm{gen}}^{-1}] =\frac{1}{4}\left(2+e^{-2r}+z+\frac{1}{z}+e^{2r}\left(1+\frac{4z}{e^{4r}+z^2}\right)\right) \, .
\label{generaldyne_bound}
\end{equation}
For $z=1$ we obtain the heterodyne detection precision bound
\begin{equation}
\Tr[\bm{F}_{\textrm{het}}^{-1}]= \frac{2\cosh^4{r}}{\cosh{(2r)}} \, ,
\label{het_bound}
\end{equation}
which coincides with the HCRB just in the $r\rightarrow\infty$ limit, thus making the heterodyne detection nearly optimal in the large squeezing regime.
However, we also point out that as the squeezing parameter grows, the overall estimation precision gets worse.
We want to investigate whether Eq.~\eqref{generaldyne_bound} is equal to the HCRB Eq.~\eqref{holevo_bound_1_mode} for some values of $z$ and $r$. 
This is equivalent to finding the zeros of the positive definite function
\begin{equation}
f(z,r) = \Tr[\bm{F}_{\textrm{gen}}^{-1}]-C^H  
\, .
\label{function_difference}
\end{equation} 
By resorting to numerical methods, we discover that the equation $f(z,r)=0$ does not have any real solutions. Therefore, we conclude that the optimal measurement must involve some non-Gaussian features, such as photon counting, as it is often the case in several other single-parameter estimation problems with Gaussian quantum statistical models~\cite{Oh2019}.
Despite this, we may still ask ourselves what is the best performance we can achieve with Gaussian resources only. This translates to finding the value of $z$ that minimizes $f(z,r)$ at fixed $r$.
We call this value $z_{opt}(r)$ and display its behavior in Fig.\eqref{optimal_z}. 
We notice that, in the absence of squeezing, the heterodyne detection scheme is the best one amongst general-dyne measurements, i.e. $z_{opt}(r=0)=1$. 
As $r$ increases, at first $z_{opt}$ decreases until it reaches $z\sim 0.6$, and then grows monotonically to asymptotically reach $1$ again. 
We also point out that, as $r$ grows, the minimum of $f(z,r)$, i.e. $f(z_{opt}(r),r)\equiv f_{opt}(r)$, decreases monotonically and approaches zero in the $r\rightarrow\infty$ limit, consistently with our previous findings, as shown in Fig.~\eqref{optimal_f}.
\section{Multiparameter quantum estimation of displaced two-mode squeezed vacuum states}
\label{s:two-mode-section}
We now consider the two-mode quantum statistical model described by displaced two-mode squeezed vacuum states, where the displacement operator to one mode only:
\begin{equation}
\ket{\psi_{\bm{\theta}}} = \left( \hat{D}(\alpha) \otimes \hat{\mathcal{I}} \right) \hat{S}^{(2)}(r)\ket{0} \, ,
\label{stat_model_2}
\end{equation}
where $\hat{D}(\alpha)$ is a displacement operator acting on the first mode, $\hat{S}^{(2)}(r)=e^{r(\hat{a}^\dagger \hat{b}^\dagger -\hat{a}\hat{b})}$ is the two-mode squeezing operator with a real squeezing parameter $r$, and $\hat{a}$ and $\hat{b}$ denote the annihilation operators of the first and second mode, respectively. 
The HCRB for the estimation of the displacement parameters only have been derived analytically in Ref.~\cite{Bradshaw2017a}, and via semi-definite programming in the case of mixed-state quantum statistical model in Ref.~\cite{Bradshaw2017}. 
Here we are interested in jointly estimating both the displacement and squeezing, hence the vector of parameters $\bm{\theta}$ is still given by Eq.~\eqref{vector_of_parameters}.
It is possible to manipulate the statistical model in order to make the calculations easier and physical interpretation of the problem clearer. In fact, the two-mode state Eq.~\eqref{stat_model_2} can actually be transformed into a tensor product of single-mode states by means of a balanced beam splitter. The justification to perform this unitary transformation lies in the fact that we can always think of the latter as part of the detection scheme. We remind the reader that the unitary evolution associated with a balanced beam splitter, i.e. a beam splitter with trasmissivity $T=1/2$, is given by 
\begin{equation}
\hat{U}_\textrm{BS}=e^{\frac{\pi}{4}(\hat{a}^\dagger \hat{b} -\hat{a}\hat{b}^\dagger)} \, .
\label{BS_unitary}
\end{equation}
The evolution of the bosonic operators under the action of Eq.~(\ref{BS_unitary}) is given by
\begin{equation}
\hat{U}^\dagger_\textrm{BS} \, \hat{a}\, \hat{U}_\textrm{BS} = \frac{\hat{a}+\hat{b}}{\sqrt{2}}, \quad \hat{U}^\dagger_\textrm{BS}\, \hat{b} \, \hat{U}_\textrm{BS} = \frac{\hat{b}-\hat{a}}{\sqrt{2}} \, .
\end{equation}
Hence, the evolution of the state in Eq.~\eqref{stat_model_2} under $\hat{U}_\textrm{BS}$ reads
\begin{equation}
\hat{U}_\textrm{BS}\ket{\psi_{\bm{\theta}}}   = 
 \hat{D}( \tfrac{\alpha}{\sqrt{2}})\hat{S}(r)\ket{0}\otimes  \hat{D}( \tfrac{-\alpha}{\sqrt{2}})\hat{S}(-r)\ket{0} \, .
\label{stat_model_2b}
\end{equation}
As the ultimate bound on the estimation precision cannot depend on the application of unitary operation on the original statistical model $\ket{\psi_{\bm{\theta}}}$, we will now consider Eq.~(\ref{stat_model_2b}) to be our new statistical model and, with a slight abuse of notation, we will still denote it with $\ket{\psi_{\bm{\theta}}}$. 
We also introduce the following basis for the two single-mode Hilbert spaces
\begin{align}
\ket{e_n} &= \hat{D}(\tfrac{\alpha}{\sqrt{2}}) \hat{S}(r)\ket{n}  \, ,\\
 \ket{f_n} &= \hat{D}(\tfrac{-\alpha}{\sqrt{2}}) \hat{S}(-r)\ket{n}  \, .
\end{align}
Hence the statistical model simply reads $\ket{\psi_{\bm{\theta}}} = \ket{e_0}\otimes\ket{f_0} \equiv \ket{e_0 f_0}$.
We can now compute the SLD-CRB, the HCRB, and discuss their experimental attainability.
In Appendix \ref{a:two-mode-calculation} we outline the detailed calculation of the SLDs and the QFI matrix. After some calculations, one finds 
\begin{equation}
\bm{Q} = \mqty (4\cosh{(2r)} & 0  & 0 \\ 0 & 4\cosh{(2r)} & 0  \\ 0 & 0 & 4) \, .
\end{equation}
As expected from symmetry arguments, similarly to the single-mode case, the QFI matrix does not depend on the displacement $\alpha$. The SLD-CRB is then given by
\begin{equation}
C^S=\Tr[\bm{Q}^{-1}]= \frac{1}{4}+\frac{1}{2\cosh{(2r)}} \, .
\label{CRB_2_mode}
\end{equation}
Contrary to the single-mode scenario, in this setting squeezing is a useful resource for our estimation protocol, i.e. the bigger the squeezing parameter, the better the achievable overall estimation precision.
Note that this is already true when we are interested in the estimation of the displacement only, as can be seen by the fact that $\bm{Q}_{33}$ does not depend on $r$. 
We can understand this behaviour intuitively by giving a closer look at our statistical model Eq.~\eqref{stat_model_2b}.
Fist of all, we notice that both modes carry complete information about the displacement $\alpha$, encoded in the unitary operators $\hat{D}(\frac{\pm\alpha}{\sqrt{2}})$. We also know from the previous section that single-mode squeezing allows for the estimation of the real (imaginary) part of $\alpha$ with arbitrary precision, in the $r\rightarrow -\infty$ ($r\rightarrow + \infty$) limit, at the cost of ignoring the imaginary (real) part altogether. 
However, Eq.~\eqref{stat_model_2b} is factorized into two single-mode states which are squeezed in orthogonal directions in the phase space. We can thus exploit structure this to optimally estimate $\theta_1=\Re{\alpha}$ and $\theta_2=\Im{\alpha}$ simultaneously by performing proper quadrature measurements on the two modes.
The Uhlmann curvature $\bm{D}$ reads
\begin{equation}
\bm{D} = \mqty (0 & 4 & 0 \\ -4 & 0 & 0 \\ 0 & 0 & 0) \, ,
\end{equation}
which is equal to that of the single-mode scenario. Therefore the quantumness of the statistical model is $R=1$, i.e. the maximum degree of incompatibility between the parameters. We suspect that this is due to the fact that we are dealing with a system at zero temperature, as seen in previous sections.

\begin{figure}
\centering 
\includegraphics[scale=0.44]{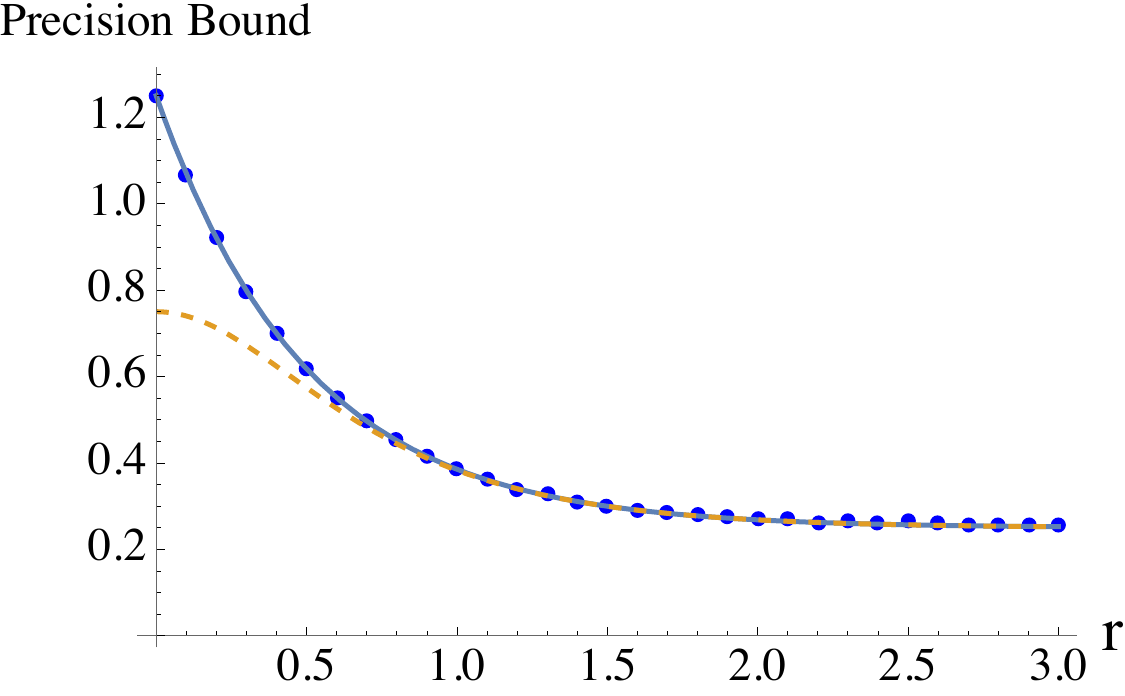}
\caption{The dashed orange line denotes the SLD-CRB Eq.~\eqref{CRB_2_mode}, while the solid blue line  denotes the estimation precision achieved with the double-homodyne detection scheme Eq.~(\ref{dual_homodyne_precision}). The dots are the result of the numerical optimization that yields the HCRB. The plot shows that $C^H$ coincides with the estimation precision obtained with the double-homodyne detection scheme, thus making the latter the optimal measurement for the joint estimation problem at study.}
\label{dual_mode}
\end{figure}

Following the techniques introduced in the previous section for the single-mode case, we know that in order to compute the HCRB it is sufficient to restrict the study of the Hermitian operators $\hat{\bm{X}} = \lbrace \hat{X}_1,\hat{X}_2,\hat{X}_3\rbrace$ satisfying the locally unbiased conditions Eq.~\eqref{locally_unbiased_cond1} and Eq.~\eqref{locally_unbiased_condition2_v2}, to the Hilbert subspace spanned by the partial derivatives of the statistical model. 
We compute these in Appendix \ref{a:two-mode-calculation}, i.e. Equations~(\ref{D1}-\ref{D3}), hence it is clear that the Hilbert subspace of interest is spanned by 
\begin{equation}
\begin{split}
    \lbrace \ket{e_0 f_0 }&  ,  \ket{e_0 f_1},\ket{e_1 f_0}, \ket{e_0 f_2}, \ket{e_2 f_0} \rbrace \equiv \\ &  \lbrace\ket{\lambda_1},\ket{\lambda_2},\ket{\lambda_3},\ket{\lambda_4},\ket{\lambda_5}\rbrace \, .
\end{split}
\end{equation}
With this new notation the statistical model reads $\ket{\psi_{\bm{\theta}}}=\ket{\lambda_1}$, and its partial derivatives are given by 
\begin{align}
\ket{\partial_1 \psi_{\bm{\theta}}} &= \frac{e^{-r}}{\sqrt{2}}\ket{\lambda_3} - i\theta_2 \ket{\lambda_1} - \frac{e^r}{\sqrt{2}} \ket{\lambda_2} \,, \\
\ket{\partial_2 \psi_{\bm{\theta}}} &= \frac{ie^{r}}{\sqrt{2}}\ket{\lambda_3} + i\theta_1 \ket{\lambda_1} - \frac{ie^{-r}}{\sqrt{2}} \ket{\lambda_2} \,,\\
\ket{\partial_3 \psi_{\bm{\theta}}} &= \frac{1}{\sqrt{2}}\ket{\lambda_5} - \frac{1}{\sqrt{2}}\ket{\lambda_4} \, .
\end{align}
The constraint imposed by Eq.~(\ref{locally_unbiased_cond1}) implies
\begin{equation}
\bra{\lambda_1} \hat{X}_1 \ket{\lambda_1}=\bra{\lambda_1} \hat{X}_2 \ket{\lambda_1}=\bra{\lambda_1} \hat{X}_3 \ket{\lambda_1} = 0\, .
\end{equation}
In Appendix \ref{a:HCRB-two-mode} we display the additional constraints on each Hermitian operator $\hat{X}_k$ imposed by Eq.~\eqref{locally_unbiased_condition2_v2}.
We end up with the vector of operators $\hat{\bm{X}}$ parameterized by fifteen free real variables.
One can then work out the three by three Hermitian matrix $\bm{Z}_{\bm{\theta}}[\hat{\bm{X}}]$
and use the latter to compute the function $h_{\bm{\theta}}[\hat{\bm{X}}]$, whose minimization yields the HCRB. We do not report the analytic expression of $h_{\bm{\theta}}[\hat{\bm{X}}]$ for brevity and also because the optimization is now non-trivial and must be carried out with numerical methods.
Before displaying the results of such minimization, we want to further discuss how to exploit the fact that the statistical model can be expressed - by means of a balanced beam-splitter - as a tensor product of two single-mode states squeezed in orthogonal directions in the phase space. 
As we have hinted before, this suggests that performing quadrature measurements on the two modes would allow us to optimally estimate the displacement.
In fact, it has been shown in Ref.~\cite{Bradshaw2017} that a double-homodyne measurement achieves optimality whenever the quantum state is entangled, if we are interested in the estimation of the displacement $\alpha$ only. Hence, it seems reasonable to choose this detection scheme as the starting point of our analysis for the three-parameter estimation case. 
The covariance matrix $\bm{\sigma}$ of the statistical model Eq.~\eqref{stat_model_2b} is simply the direct sum of the covariance matrices associated with single-mode squeezed vacuum states, namely
\begin{equation}
\bm{\sigma}=\text{diag}(e^{2r},e^{-2r},e^{-2r},e^{2r}) ,
\end{equation}
while the statistical model's vector of first moments reads
\begin{equation}
\overline{\bm{r}} = (\theta_1,\theta_2,-\theta_1,-\theta_2)^\intercal \, .
\end{equation}
The double-homodyne detection scheme consists in performing a $\hat{p}$ quadrature measurement on the first mode, and a $\hat{q}$ quadrature measurement on the second one (we should remind ourselves that these measurements are performed after a balanced beam-splitter, if we consider the state in Eq.~\eqref{stat_model_2}) as our initial state).
Consequently, the covariance matrix $\bm{\sigma}_m$ that characterizes this POVM reads 
\begin{equation}
 \bm{\sigma}_m = \lim_{z\rightarrow 0}\text{diag}(z^{-1},z,z,z^{-1}) \, .
\end{equation}
Hence, the corresponding conditional probability distribution of outcomes is a Gaussian centered in $\overline{\bm x}=\overline{\bm r}$, with covariance matrix given by $\boldsymbol{\Sigma}=(\boldsymbol{\sigma}+\boldsymbol{\sigma}_m)/2$.
By taking the limit $z\to 0$ after the matrix inversion, one can prove that
\begin{equation}
(\bm{\sigma} + \bm{\sigma}_m)^{-1} =
\text{diag}(0,e^{2r},e^{2r},0) \, .
\end{equation}
The FI matrix can then be evaluated using Eq.~(\ref{FImatrixGauss}), yielding
\begin{equation}
\bm{F} = \mqty (2e^{2r} & 0 &  0 \\ 0 & 2e^{2r} & 0 \\ 0  & 0 & 4) \, .
\end{equation}
The corresponding estimation precision of the double-homodyne detection scheme thus reads
\begin{equation}
\Tr[\bm{F}^{-1}] =\frac{1}{4}+e^{-2r} \, .
\label{dual_homodyne_precision}
\end{equation}
It turns out that this bound coincides with the HCRB we obtained from numerical optimization, as it can be seen in Fig.~\eqref{dual_mode}.
Hence, we may infer that the expression of the HCRB is given by
\begin{equation}
C^H = \frac{1}{4} + e^{-2r}
\end{equation}
and that the double-homodyne detection scheme represents the optimal POVM for our estimation problem, for every value $r$. Also in this case, the AI measure $R$ largely overestimates the true gap between the HCRB and the SLD-CRB bound.
We recall that, given the statistical model Eq.~(\ref{stat_model_2b}), the double-homodyne represents the optimal measurement for the estimation of the displacement only. Here we have shown that from the measurement outcomes of that same POVM one can \emph{optimally} estimate the squeezing parameter as well, without incurring in any additional cost.

\section{Conclusions}
\label{s:conclusions}
We have analyzed in details three-parameter estimation problems involving pure Gaussian states, focusing on the joint estimation of complex displacement and real squeezing, encoded unitarily in single- and two-mode pure states. We have explicitly evaluated the Holevo-Cramér-Rao bound (HCRB) and explored its dependence on the degree of squeezing.

Our findings reveal that for single-mode states, the HCRB deteriorates as squeezing increases. In scenarios of substantial squeezing, heterodyne detection approaches near-optimality. However, our analysis reveals that, more broadly, achieving the optimal measurement requires non-Gaussian measurements. Conversely, in the two-mode scenario, we discover that squeezing—or equivalently, entanglement—serves to enhance the HCRB. Remarkably, this enhanced bound is attainable through double-homodyne detection across all levels of squeezing, showcasing the intrinsic value of entanglement in quantum metrology.

Our results reaffirms the role of entanglement as a resource in quantum metrology, setting the stage for advanced quantum sensing technologies. By clarifying the  relationship between squeezing, measurement optimality, and quantum state estimation, we pave the way to further improve metrological precision with realistic and reliable detection schemes.

\section{Acknowledgments}
MGG acknowledges useful discussions with Francesco Albarelli. GB is part of the AppQInfo MSCA ITN which received funding from the European Union’s Horizon 2020 research and innovation programme under the Marie Sklodowska-Curie grant agreement No 956071. MGG has been partially supported by MUR through Project No. PRIN22-2022KB2JJM-CONTRABASS. The work of MGAP has been done under the auspices of 
GNFM-INdAM and has been partially supported by MUR through Project No. PRIN22-2022T25TR3-RISQUE.

\bibliographystyle{unsrt}
\bibliography{pgHCRB3Notes}

\appendix

\section{QFI matrix evaluation for single-mode Gaussian quantum state}
\label{a:single-mode-calculations}
In this appendix we derive the QFI matrix for the single-mode quantum statistical model Eq.~\eqref{stat_model_1}.
We remind the reader that for pure statistical models the SLDs can be expressed as in Eq.~\eqref{SLDpure}.
Hence, in order to compute $\hat{L}_{\mu}$ we just need to differentiate the statistical mode with respect to the parameter to estimated $\theta_\mu$.
To this end we first compute the derivatives of the displacement operator and the squeezing operator with respect to the parameters to estimate, namely
\begin{align}
\partial_{1} \hat{D}(\alpha) &= (\hat{a}^\dagger - \hat{a}+i\theta_2)\hat{D}(\alpha) \,, \\
\partial_{2} \hat{D}(\alpha) &= i(\hat{a}^\dagger + \hat{a}-\theta_1)\hat{D}(\alpha) \,, \\
\partial_{3} \hat{S}(r) &= \frac{\hat{a}^{\dagger 2}-\hat{a}^2}{2}\hat{S}(r) \, .
\label{derivative_squeezing} 
\end{align}
Exploiting these and the identities $\hat{D}^\dagger(\alpha) \hat{a} \hat{D}(\alpha) = \hat{a}+\alpha$ and $\hat{S}^\dagger (r) \hat{a} \hat{S}(r)=\mu \hat{a} + \nu \hat{a}^\dagger$, where $\mu=\cosh{r}$ and $\nu=\sinh{r}$, we finally obtain
\begin{align}
\ket{\partial_1\psi_{\bm{\theta}}}  &= -i\theta_2 \ket{e_0} + e^{-r}\ket{e_1} \, , 
\label{d1} \\
\ket{\partial_2\psi_{\bm{\theta}}} &= i\theta_1 \ket{e_0} + ie^r \ket{e_1} \, ,
\label{d2} \\
\ket{\partial_3\psi_{\bm{\theta}}} &=\frac{1}{\sqrt{2}}\ket{e_2} \, .
\label{d3}
\end{align}
Substituting these expressions into Eq.~\eqref{SLDpure} yields the SLDs, i.e. 
\begin{align}
\hat{L}_1 &= 2e^{-r} \left( \ket{e_1}\!\!\bra{e_0} + \ket{e_0}\!\!\bra{e_1} \right) \,,\\
\hat{L}_2 &= 2ie^{r} \left(\ket{e_1}\!\!\bra{e_0} - \ket{e_0}\!\!\bra{e_1}\right) \,, \\
\hat{L}_3 &= \sqrt{2} \left(\ket{e_2}\!\!\bra{e_0} + \ket{e_0}\!\!\bra{e_2}\right) \,.
\end{align}
We can then evaluate the QFI matrix, whose expression for our pure statistical model $\ket{\psi_{\bm{\theta}}}=\ket{e_0}$ reads
\begin{equation}
\bm{Q}_{\mu\nu} = \frac{1}{2}\bra{e_0} \hat{L}_\mu \hat{L}_\nu + \hat{L}_\nu \hat{L}_\mu \ket{e_0} \, .
\end{equation}
After some calculations we obtain the following diagonal matrix
\begin{equation}
\bm{Q} = \mqty (4e^{-2r} & 0  & 0 \\ 0 & 4e^{2r} & 0  \\ 0 & 0 & 2) \, .
\end{equation}

\section{QFI matrix evaluation for single-mode Gaussian quantum state}
\label{a:two-mode-calculation}
In this appendix we derive the QFI matrix for the two-mode quantum statistical model in Eq.~(\ref{stat_model_2b}).
Let us first work out the derivatives of $\hat{D}(\tfrac{\pm\alpha}{\sqrt{2}})$ and $\hat{S}(\pm r)$ with respect to the parameters to be estimated $\theta_{\mu}$, namely
\begin{align}
\partial_{1}\hat{D}(\tfrac{\pm\alpha}{\sqrt{2}}) &= \left(\frac{\pm \hat{a}^\dagger\mp \hat{a}}{\sqrt{2}}+\frac{i\theta_2}{2}\right)\hat{D}(\tfrac{\pm\alpha}{\sqrt{2}}) \,, \\
\partial_{2}\hat{D}(\tfrac{\pm\alpha}{\sqrt{2}})  &= i \left(\pm\frac{\hat{a}+\hat{a}^\dagger}{\sqrt{2}}-\frac{\theta_1}{2}\right)\hat{D}(\tfrac{\pm\alpha}{\sqrt{2}}) \,, \\
\partial_{3}\hat{S}(\pm r) &= \pm\left(\frac{\hat{a}^{\dagger 2}-\hat{a}^2}{2} \right) \hat{S}(\pm r) \, .
\end{align}
Using these expressions, together with the previously introduced transformation properties of the mode operators under displacement and local squeezing unitaries, we cam compute
the derivative of our statistical model.  After cumbersome calculations one obtains
\begin{align}
\ket{\partial_1\psi_{\bm{\theta}}} &= \frac{e^{-r}}{\sqrt{2}}\ket{e_1 f_0}  - i\theta_2 \ket{e_0 f_0} - \frac{e^r}{\sqrt{2}} \ket{e_0 f_1} \, , 
\label{D1} \\
\ket{\partial_2\psi_{\bm{\theta}}} &= \frac{ie^r}{\sqrt{2}} \ket{e_1 f_0} + i\theta_1 \ket{e_0 f_0} -\frac{ie^{-r}}{\sqrt{2}}\ket{e_0 f_1} \, ,
\label{D2} \\
\ket{\partial_3\psi_{\bm{\theta}}} 
&=\frac{1}{\sqrt{2}}\ket{e_2 f_0} - \frac{1}{\sqrt{2}} \ket{e_0 f_2} \, .
\label{D3}
\end{align}
Substituting these expressions into Eq.~\eqref{SLDpure} yields the SLDs
\begin{align}
\hat{L}_1 &= \sqrt{2} \left( e^{-r} \ket{e_1 f_0}\!\!\bra{e_0 f_0} - e^{r} \ket{e_0 f_1}\!\!\bra{e_0 f_0} + h.c. \right) \, ,
\\
\hat{L}_2 &= i \sqrt{2} \left( e^{r} \ket{e_1 f_0}\!\!\bra{e_0 f_0} - e^{-r} \ket{e_0 f_1}\!\!\bra{e_0 f_0} - h.c.  \right) \, ,
\\
\hat{L}_3 &= \sqrt{2} \left( \ket{e_2 f_0}\!\!\bra{e_0 f_0} - \ket{e_0 f_2}\!\!\bra{e_0 f_0} + h.c. \right) \, .
\end{align}
We can then evaluate the QFI matrix $\bm{Q}$, whose matrix elements are given by 
\begin{equation}
\bm{Q}_{\mu\nu} = \frac{1}{2}\bra{e_0 f_0 } \hat{L}_\mu \hat{L}_\nu + \hat{L}_\nu \hat{L}_\mu \ket{e_0 f_0 } \, .
\end{equation}
After some calculations we obtain the following diagonal matrix 
\begin{equation}
\bm{Q} = \mqty (4\cosh{(2r)} & 0  & 0 \\ 0 & 4\cosh{(2r)} & 0  \\ 0 & 0 & 4) \, .
\end{equation}

\section{HCRB evaluation for two-mode displaced squeezed vacuum states}
\label{a:HCRB-two-mode}
In this appendix we outline the computation of the HCRB for the two-mode quantum statistical model Eq.~\eqref{stat_model_2b}.
The locally unbiased condition Eq.~\eqref{locally_unbiased_condition2_v2} imposes three additional constraints for each Hermitian operator $\hat{X}_\mu$.
In particular, one finds that $\hat{X}_1$ must satisfy
\begin{align}
&\sqrt{2} \left[ e^{-r}\Re{\bra{\lambda_1}\hat{X}_1\ket{\lambda_3}}-e^r\Re{\bra{\lambda_1}\hat{X}_1\ket{\lambda_2}} \right] = 1 \,, \nonumber \\
&-\sqrt{2} \left[ e^{r}\Im{\bra{\lambda_1}\hat{X}_1\ket{\lambda_3}}-e^{-r}\Im{\bra{\lambda_1}\hat{X}_1\ket{\lambda_2}} \right] = 0 \,, \nonumber \\
&\sqrt{2} \left[ \Re{\bra{\lambda_1}\hat{X}_1\ket{\lambda_5}} - \Re{\bra{\lambda_1}\hat{X}_1\ket{\lambda_4}} \right] = 0 \, . \nonumber
\end{align}
These conditions in turn imply that
\begin{align}
\bra{\lambda_1}\hat{X}_1\ket{\lambda_2} &= x_1+ie^{2r}x_2\, , \nonumber \\
\bra{\lambda_1}\hat{X}_1\ket{\lambda_3} &= \frac{e^r}{\sqrt{2}}+e^{2r} x_1+ ix_2 \, , \nonumber \\
\bra{\lambda_1}\hat{X}_1\ket{\lambda_4} &= x_3+ix_4 \,,\nonumber \\
\bra{\lambda_1}\hat{X}_1\ket{\lambda_5} &= x_3+ix_5 \, , \nonumber
\end{align}
where $\lbrace x_1,x_2,x_3,x_4,x_5 \rbrace$  are five free real parameters. 
Analogously, the additional constraints on $\hat{X}_2$ read
\begin{align}
&\sqrt{2} \left[ e^{-r}\Re{\bra{\lambda_1}\hat{X}_2\ket{\lambda_3}}-e^r\Re{\bra{\lambda_1}\hat{X}_2\ket{\lambda_2}} \right] = 0 \, , \nonumber \\
&-\sqrt{2} \left[ e^{r}\Im{\bra{\lambda_1}\hat{X}_2\ket{\lambda_3}}-e^{-r}\Im{\bra{\lambda_1}\hat{X}_2\ket{\lambda_2}} \right] = 1 \, , \nonumber \\
&\sqrt{2} \left[ \Re{\bra{\lambda_1}\hat{X}_2\ket{\lambda_5}} - \Re{\bra{\lambda_1}\hat{X}_2\ket{\lambda_4}} \right] = 0 \, . \nonumber
\end{align}
These are equivalent to
\begin{align}
\bra{\lambda_1}\hat{X}_2\ket{\lambda_2} &= x_6+i\left( \frac{e^r}{\sqrt{2}} + e^{2r}x_6	\right) \,, \nonumber  \\
\bra{\lambda_1}\hat{X}_2\ket{\lambda_3} &= e^{2r} x_6+ ix_7 \, , \nonumber \\
\bra{\lambda_1}\hat{X}_2\ket{\lambda_4} &= x_8+ix_9 \,,  \nonumber \\
\bra{\lambda_1}\hat{X}_2\ket{\lambda_5} &= x_8+ix_{10} \, , \nonumber
\end{align}
with $\lbrace x_6,x_7,x_8,x_9,x_{10} \rbrace$ five other free real parameters.
Finally, the constraints on $\hat{X}_3$ are
\begin{align}
&\sqrt{2} \left[ e^{-r}\Re{\bra{\lambda_1}\hat{X}_3\ket{\lambda_3}}-e^r\Re{\bra{\lambda_1}\hat{X}_3\ket{\lambda_2}} \right] = 0 \, , \nonumber  \\
&-\sqrt{2} \left[ e^{r}\Im{\bra{\lambda_1}\hat{X}_3\ket{\lambda_3}}-e^{-r}\Im{\bra{\lambda_1}\hat{X}_3\ket{\lambda_2}} \right] = 0 \, , \nonumber \\
& \sqrt{2} \left[ \Re{\bra{\lambda_1}\hat{X}_3\ket{\lambda_5}} - \Re{\bra{\lambda_1}\hat{X}_3\ket{\lambda_4}} \right] = 1 \, . \nonumber
\end{align}
Hence, the matrix elements of $\hat{X}_3$ must satisfy
\begin{align}
\bra{\lambda_1}\hat{X}_3\ket{\lambda_2} & =x_{11} + ie^{2r}x_{12}\, , \nonumber \\
\bra{\lambda_1}\hat{X}_3\ket{\lambda_3} &= e^{2r}x_{11} + ix_{12} \, , \nonumber \\
\bra{\lambda_1}\hat{X}_3\ket{\lambda_4}& = x_{13} + ix_{14} \, , \nonumber \\ 
\bra{\lambda_1}\hat{X}_3\ket{\lambda_5} &= x_{13} + \frac{1}{\sqrt{2}}+ix_{15}\, , \nonumber
\end{align}
where $\lbrace x_{11},x_{12},x_{13},x_{14},x_{15}\rbrace$ are free real parameters.
The hermiticity of these operators imposes additional constraints on the matrix elements and, as explained Section~\ref{s:single-mode-section} of the main text, we can set to zero all other matrix elements of $\hat{X}_\mu$ not involved in the above-mentioned constraints. We therefore have a total of fifteen free real variables.

\end{document}